\documentclass[%
 reprint,
 amsmath,amssymb,
 aps,
]{revtex4-2}

\usepackage{graphicx}
\usepackage{dcolumn}
\usepackage{bm}
\usepackage{makecell} 
\usepackage{tikz}
\usepackage{subcaption}

\usetikzlibrary{arrows.meta}
 \usetikzlibrary{calc}

\begin{document}

\preprint{APS/123-QED}

\title{Achromatic Telescopic Squeezing for Dynamic Aperture Optimization in the Electron Storage Ring of the EIC}

\author{Jonathan Unger}
 \email{jeu8@cornell.edu}

\author{Georg Hoffstaetter}%
 
\affiliation{%
 Cornell University, Ithaca, New York 14850, USA
}%

\date{\today}

\begin{abstract}
We investigate the application of the Achromatic Telescopic Squeezing (ATS) scheme to the Electron Storage Ring (ESR) of the Electron-Ion Collider (EIC) as a method to improve dynamic aperture and momentum acceptance. A comparative study is performed between conventional sextupole correction schemes and ATS-based optics using both a simplified test lattice and the full ESR lattice. We show that ATS optics can reduce the required sextupole strengths and mitigate higher-order nonlinear effects, leading to improved momentum aperture. With the ATS principle one could reduce the number of sextupoles in an arcs by a factor of two while maintaining similar momentum aperture. We additionally show a scheme utilizing all sextupoles which provides an advantage in momentum aperture. While the resulting ATS optics provides a measurable increase in momentum acceptance ($\sim$0.1\% under the test conditions), it also induces emittance growth due to increased $\beta$-functions in the arcs. This trade-off limits its applicability for the ESR but suggests potential advantages for storage rings where moderate emittance growth is acceptable.

\end{abstract}

\maketitle

\section{introduction}

The Electron-Ion Collider (EIC) \cite{CDR}, planned for construction at Brookhaven National Laboratory, will enable high-luminosity collisions of polarized electron and ion beams over a range of energies. Achieving the required luminosity requires strong focusing at the interaction points (IPs), producing nonlinear beam dynamics which present difficulties in maintaining sufficient transverse and momentum aperture.

Chromatic effects originating from the interaction regions (IRs) introduce significant nonlinearities that must be corrected. However, conventional sextupole correction strategies can introduce additional higher-order resonances that limit dynamic aperture. This effect becomes more severe at higher energies, where stronger focusing is needed to compensate for higher radiative emittance creation. And stronger focusing requires stronger sextupoles for chromaticity correction.

In this work, we investigate the application of the Achromatic Telescopic Squeezing (ATS) scheme, originally developed for the Large Hadron Collider \cite{PhysRevSTAB.16.111002}. While the LHC application of ATS uses beta-beats in the arcs to reduce  beam sizes at the collision points, we propose to use this beta-beat to increase the effect of sextupoles, which can then be reduced in strength. This can allow the use of weaker or fewer sextupoles and offer a potential increase to dynamic aperture.

We compare this ATS-based correction schemes with conventional two-family and four-family sextupole configurations \cite{osti_1875587} using both a simplified test lattice and the full ESR lattice.

We demonstrate that ATS can improve momentum aperture through reduced sextupole strength and through more favorable sextupole configurations. However, this improvement comes at the cost of increased emittance due to larger beta functions in the arcs. The implications of this trade-off are evaluated in the context of ESR performance requirements.

\section{Ring design}

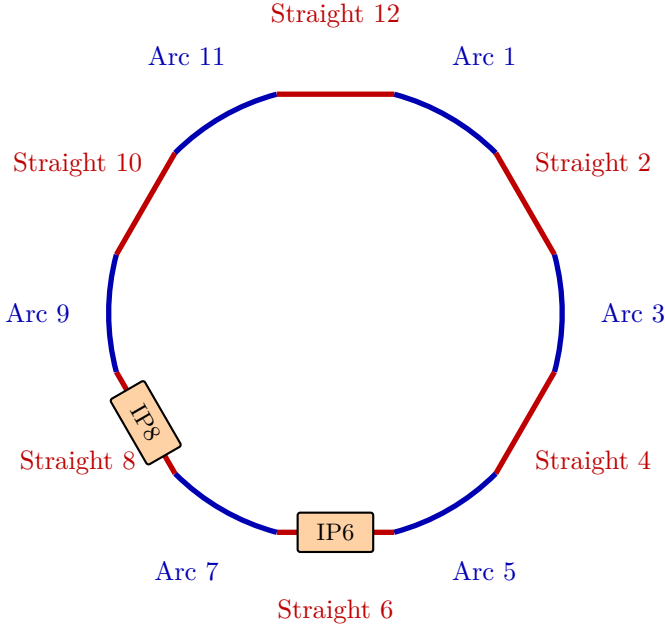
\begin{figure}[htbp]
\centering

\begin{tikzpicture}[
scale=0.75,
transform shape=false,
arc/.style={line width=2pt, blue!70!black},
straight/.style={line width=2pt, red!75!black},
seglabel/.style={
font=\normalsize,
inner sep=1pt
},
detector/.style={
draw=black,
fill=orange!35,
rounded corners=1pt,
line width=0.8pt,
minimum width=1.0cm,
minimum height=0.52cm,
font=\small
}
]

\def\R{4.0}
\def\thetaZero{75}
\def\labelR{5.25}

\foreach \n in {1,...,12} {

\pgfmathtruncatemacro{\isOdd}{mod(\n,2)}
\pgfmathsetmacro{\angA}{\thetaZero-(\n-1)*30}
\pgfmathsetmacro{\angB}{\thetaZero-\n*30}
\pgfmathsetmacro{\mid}{(\angA+\angB)/2}

\ifnum\isOdd=1

\draw[arc]
({\angA}:\R)
arc[start angle=\angA,end angle=\angB,radius=\R];

\node[seglabel,text=blue!70!black]
at ({\mid}:\labelR)
{Arc \n};

\else

\coordinate (A) at ({\angA}:\R);
\coordinate (B) at ({\angB}:\R);

\draw[straight] (A)--(B);

\node[seglabel,text=red!75!black]
at ({\mid}:\labelR)
{Straight \n};

\ifnum\n=6
\path let \p1=(A),\p2=(B) in
node[
detector,
rotate={atan2(\y2-\y1,\x2-\x1)+180}
]
at ($(A)!0.5!(B)$)
{IP6};
\fi

\ifnum\n=8
\path let \p1=(A),\p2=(B) in
node[
detector,
rotate={atan2(\y2-\y1,\x2-\x1)+180}
]
at ($(A)!0.5!(B)$)
{IP8};
\fi

\fi
}

\end{tikzpicture}

\caption{
Layout of the ESR. Placement within the RHIC tunnel requires departures
from perfect six-fold symmetry. Dynamic aperture is evaluated for both
a realistic ESR lattice and an idealized symmetric test lattice.
}

\label{ESR_layout}
\end{figure}

The ESR \cite{CDR} is comprised of six arcs and six straight sections, alternating to produce roughly circular ring. As there are twelve sections overall, the names are determined by their clock position, with straight sections being even and the arcs odd numbered. The interaction point of the ring is at 6 o'clock position with a second IR available in straight section 8, shown schematically in Fig. \ref{ESR_layout}. The ESR lattice is under development; the version 5.3 used for the present study houses the RF cavities in section 10, and the beam is injected in section 12. In order to achieve the desired longitudinal polarization at the IPs, spin rotators containing solenoids and skew quadrupoles are placed on both ends of each interaction region. In order to increase luminosity, crab cavities are included before and after each IP.

\section{Baseline Dynamic Aperture Optimization}

For the purpose of this study, dynamic aperture optimization is done for an 18GeV configuration of the ESR. The ESR will also be operated at lower energies; we study this highest energy because it tends to produce the lowest dynamics aperture. The general correction scheme used at 18GeV in both the 1-IP and 2-IP lattices for the ESR has two families of sextupoles per plane in each arc. Even though one would like to minimize nonlinear motion from sextupoles, these families cannot be avoided because they are required for the correction of the W-function and of the chromaticities in order to have stable motion over the energy spread of the beam. These corrections are often not enough for satisfactory momentum aperture, and the ESR also utilizes second order dispersion correction through additional sextupoles. The use of all these sextupoles limits the transverse dynamic aperture but increases the momentum aperture. To reduce third-order resonances, 12 harmonic sextupoles are introduced in straight section 2.
 
\begin{figure}[ht]
    \centering

    \begin{subfigure}{1\columnwidth}
        \centering
        \includegraphics[width=\linewidth]{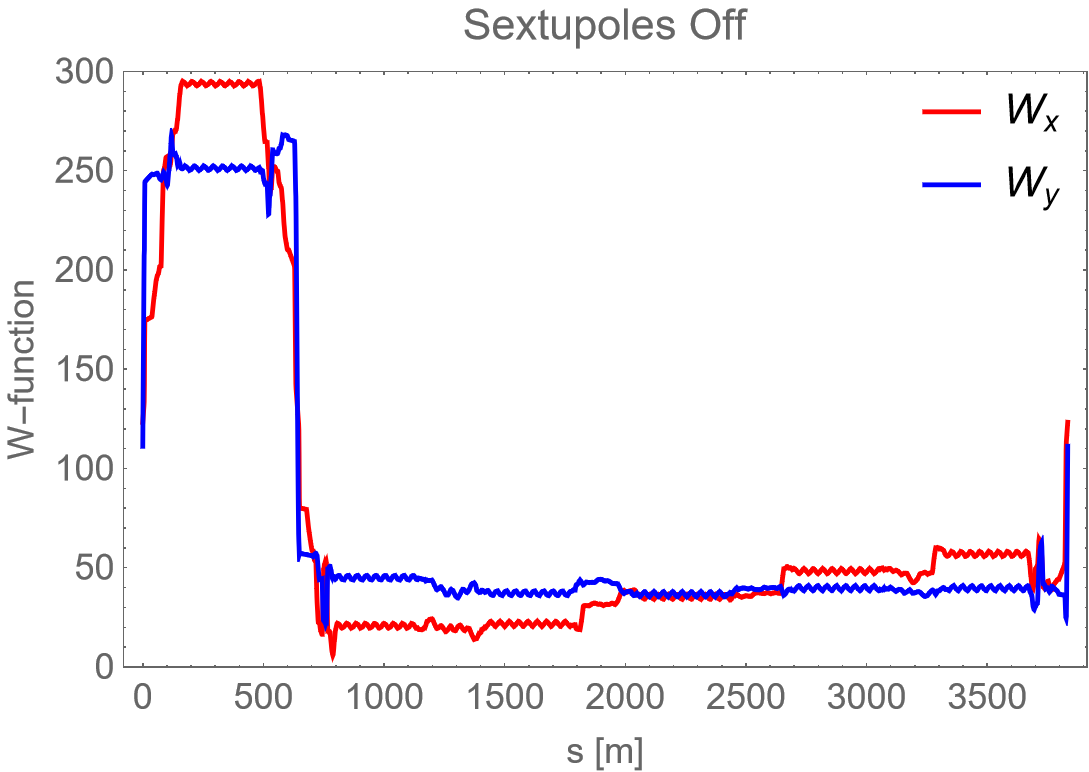}
        \caption{W-function}
        \label{fig:top}
    \end{subfigure}

    \begin{subfigure}{0.9\columnwidth}
        \centering
        \includegraphics[width=\linewidth]{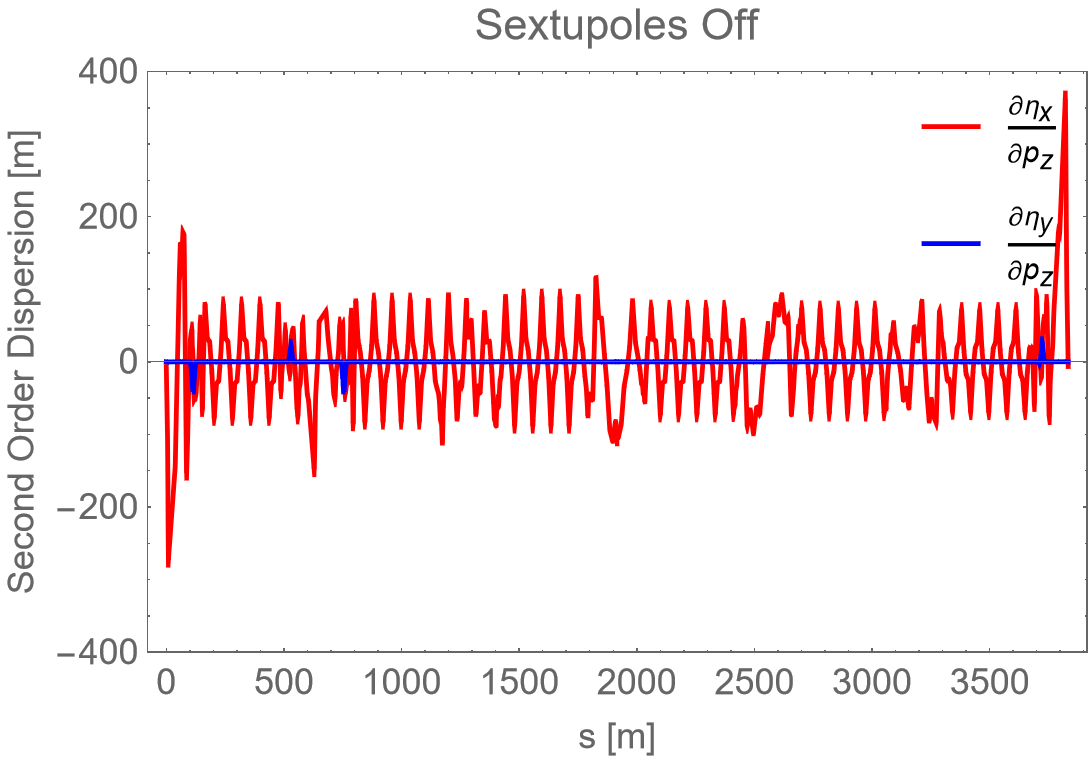}
        \caption{Second order dispersion.}
        \label{fig:bottom}
    \end{subfigure}
    \caption{Natural W-function and second order dispersion in the ESR ring with all sextupoles turned off.}
    \label{ESRWD2}
\end{figure}

\begin{figure}[ht]
    \centering
    \begin{subfigure}{1\columnwidth}
        \centering
        \includegraphics[width=\linewidth]{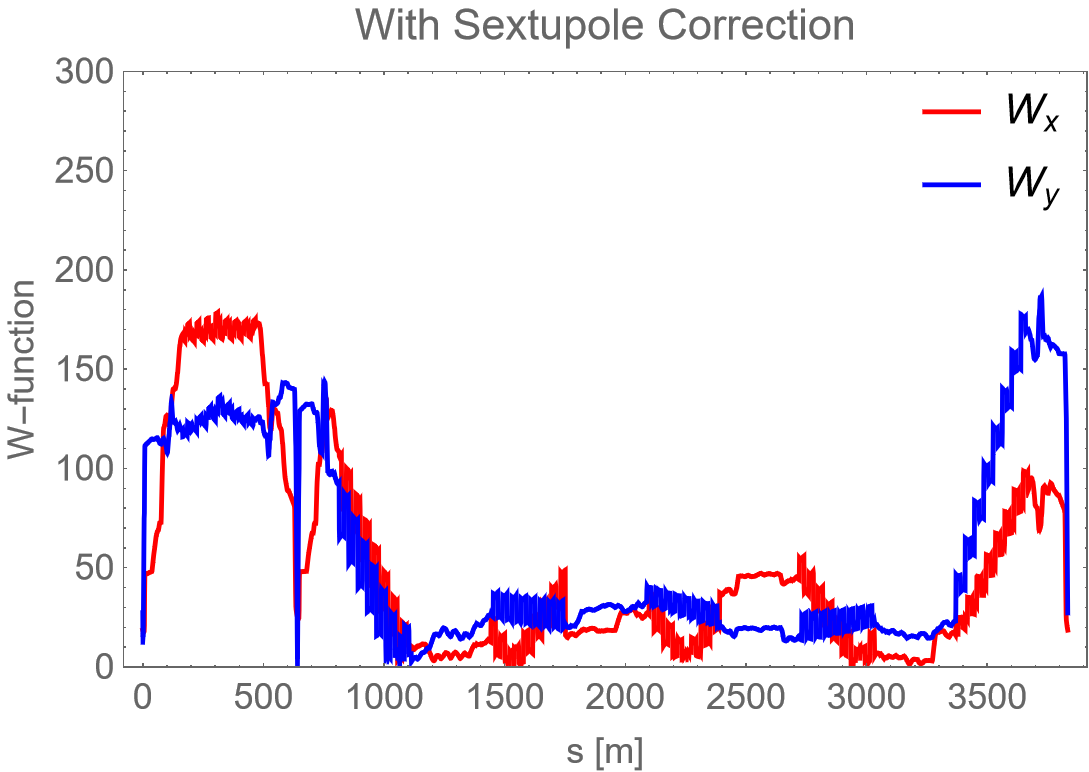}
        \caption{W-function}
        \label{fig:top}
    \end{subfigure}

    \begin{subfigure}{0.9\columnwidth}
        \centering
        \includegraphics[width=\linewidth]{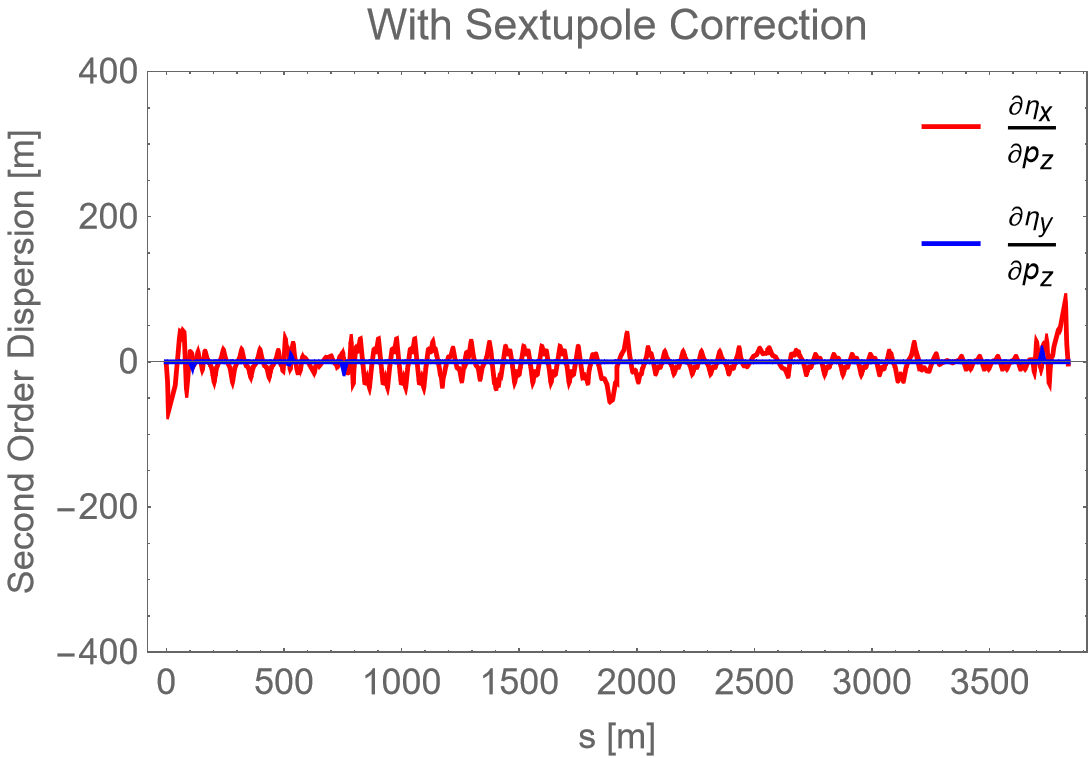}
        \caption{Second order dispersion.}
        \label{fig:bottom}
    \end{subfigure}
    \caption{W-function and second order dispersion in the ESR with 4 sextupole families to minimize chromatic functions. Values strongly reduced from Fig. \ref{ESRWD2}.}
    \label{ESRWD2Corrected}
\end{figure}

\begin{figure}[ht]
    \centering
    \includegraphics[width=1\linewidth]{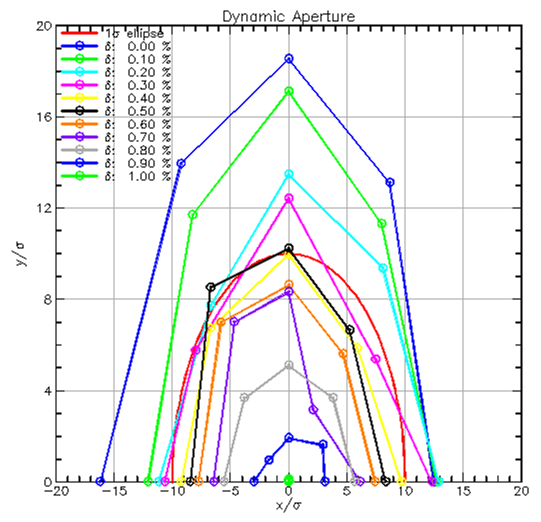}
    \caption{ESR dynamic aperture with the 4 sextupole family chromatic correction scheme with additional higher order corrections.}
    \label{ESRDA}
\end{figure}

The above scheme is not sufficient for the 2-IP case, where chromatic effects are increased by the second IR regions and where less space is available to reduce the W-function. This requires an additional correction in arc 7 to reduce the second order W-function. This is accomplished through the use of four families of sextupoles per plane while keeping the first order W-function correction in place. Fig. \ref{ESRWD2} shows the chromatic optics before this correction which is in comparison to Fig. \ref{ESRWD2Corrected} after this baseline correction. This procedure leads to the DA in Fig. \ref{ESRDA} with a sufficiently large transverse aperture and a momentum aperture just below the requirements \cite{osti_1875587}.

\section{Achromatic Telescopic Squeezing}

As the DA optimization for the ESR at 18GeV was difficult to obtain and in particular left a small margin for errors in the momentum aperture, additional schemes were explored. One promising method was Achromatic Telescopic Squeezing (ATS).

The ATS scheme was first introduced at the LHC in order to fully utilize the existing large aperture in the arcs for luminosity optimization. Because the arc has to accommodate a lower energy beam at injection, it's aperture at high energy is unnecessarily large at higher energies. The ATS scheme induces a $\beta$-beat in the arcs neighboring the IR which uses this available aperture, and this beta beat can reduce the cross section at the IP without strengthening the final focus quadrupoles. The ratio of the new peak $\beta$ in the arc to the original one is called the telescopic ratio which will be denoted by $T_r$. This value is free in the scheme which can be optimized for the lattice. A result of this, in a 90\textdegree lattice with four sextupole families, is that two of the families will be strengthened by an increase in $\beta$-function and two will be weakened by a decrease in $\beta$-function (which will be called the Strong and Weak families) \cite{PhysRevSTAB.16.111002}. This gives a different sextupole solution for the correction of chromaticities and chromatic beta beats. The different chromatic correction, although not the primary reason for the development of the ATS scheme in LHC, has possible advantages in DA optimization with potential benefit to the EIC.

Possible advantages of the ATS scheme can be explained by looking at two arguments. The first is that W-function correction is now aided by the arc quads, reducing the needed sextupole strengths. The W-function \cite{Montague:443342} is defined by

\begin{equation}
    W=\sqrt{\left(\frac{\partial\alpha}{\partial\delta}-\frac{\alpha}{\beta}\frac{\partial\beta}{\partial\delta}\right)^2+\left(\frac{1}{\beta}\frac{\partial\beta}{\partial\delta}\right)^2}
\end{equation}

And for a lattice with only quadrupoles and sextupoles, this can be approximated by the absolute value of

\begin{equation}
    \int_{S_0}^{S_0+C}ds[k_1(s)-k_2(s)D_x(s)] 
    \beta_{x,y}(s)e^{2j[\mu_x,y(s)-\mu_x,y(s_0)]}   
\end{equation}

This means that each quadrupole in a FODO arc with periodic optics will be compensated by the next quadrupole. Also, because of this phase factor in the W-function, sextupoles in neighboring cells will act counter each other. This leads to the general relation in a 90\textdegree FODO section with standard 4-family scheme that chromaticity is corrected by the average sextupole strength while W-function is corrected by the difference. Looking at the correction in just one plane with properly matched phasing, this leads to the W-function correction for IR over one arc to be   

\begin{equation}
   \frac{1}{2} N\beta\eta\Delta k_2=I_W
\end{equation}

\noindent Where $N$ is the number of cells, $\Delta k_2$ is the difference in sextupole strength between the two families, and $I_W$ is the W-function contribution from the IR. The needed sextupole strengths are then

\begin{equation}
    \Delta k_2=\frac{2I_W}{N\beta\eta}
\end{equation}

In the ATS optics, this can be improved. For now, we assume that the weak sextupoles are turned off, leading to

\begin{equation}
    \frac{1}{2}N\beta T_r\eta k_s+\frac{1}{2}N\beta \left(T_r-1/T_r\right) k_1=I_W
\end{equation}

\noindent the $\beta$-functions here are the pre $\beta$-beat values, adjusted with the telescopic ratio. Solving for sextupole strength gives

\begin{multline}
        k_s=\frac{2I_W}{N\beta T_r \eta}-\frac{k_1}{\eta}\left(1-\frac{1}{T_r^2}\right) \\ = \frac{\Delta k_2}{T_r}-\frac{k_1}{\eta}\left(1-\frac{1}{T_r^2}\right)
\end{multline}

\noindent leading to the expected reduction in sextupole strength by a factor of $T_r$ along with a second term that approaches $k_1/\eta$ for increasing $T_r$. This reduction is seen for the difference in sextupole strengths of the two families, where in the ATS one family is currently set to zero strength. The need for a non-zero second sextupole family in the common scheme is for chromaticity correction, which is given by

\begin{equation}
    N\beta\eta \overline{k_2}=I_Q
\end{equation}

Using the same arguments as for the W-function, the ATS scheme yields

\begin{equation}
    \frac{1}{2}N\beta T_r\eta k_s=I_Q
\end{equation}

This shows that the same chromaticity correction for $k_s=\overline{k_2}$ is accomplished using half the sextupoles when $T_r=2$, so additional chromaticity correction outside of the ATS arcs is not needed in comparison to the standard scheme.

The chromaticity contribution of the ATS arc quads will slightly increase over the periodic arc cell

\begin{equation}
    Q'_{arc}=\frac{1}{2}N\beta k_1\left(T_r+1/T_r\right)
\end{equation}

The second term can later be canceled with the use of the weak sextupoles while minimally impacting the W-function correction which, when paired with the telescopic ratio of 2, will remove this concern.

A second argument for the ATS scheme brings up the topic of interleaved versus non-interleaved sextupole schemes. In this study, an interleaved sextupole section refers to one of the arcs with a repeating set of unequal sextupoles strengths. A section where all sextupoles have the same strength would be non-interleaved. The arcs in the ATS scheme are not fully non-interleaved even with the weak sextupoles turned off, as the strong families will still be interleaved with the focusing and defocusing sextupole families having different strengths. This setup where all sextupoles for each plane are equal will be referred to as a partially interleaved sextupole scheme. The ATS scheme also has the benefit of reducing the effects of the remaining interleaved families due to the $\beta$-function being increased at only one family per plane.

\begin{figure*}[ht]
\centering
\begin{subfigure}{0.5\textwidth}
  \centering
  \includegraphics[width=\textwidth]{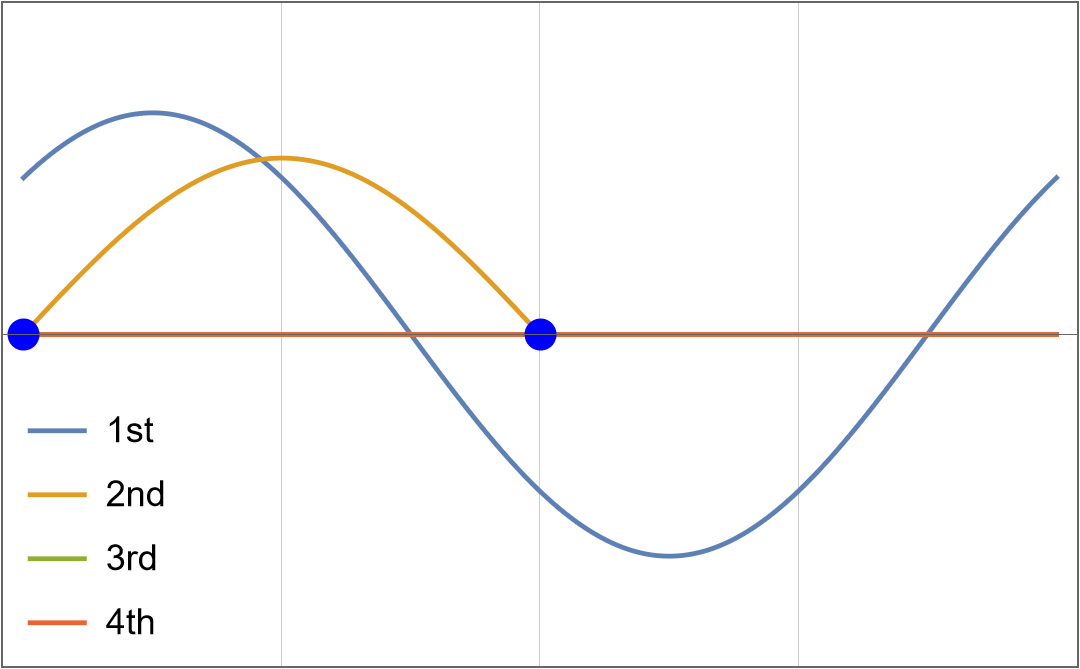}
  \caption{Non-interleaved}
  \label{fig:sub1}
\end{subfigure}%
\begin{subfigure}{0.5\textwidth}
  \centering
  \includegraphics[width=\textwidth]{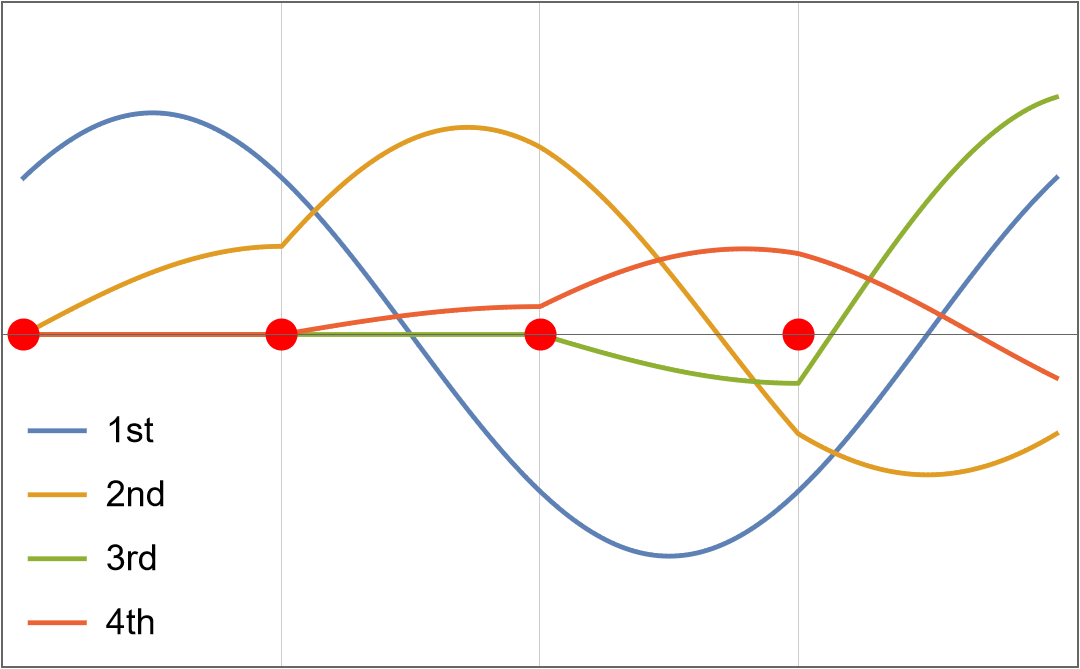}
  \caption{Interleaved}
  \label{fig:sub2}
\end{subfigure}
\caption{The advantage of non-interleaved sextupole schemes. This shows a particle traveling through a constant quadrupole field, with vertical lines showing 90\textdegree\  points. The particle is started with an initial offset and momentum, with the contributions to the orbit shown order by order. In the non-interleaved case, sextupoles are spaced 180\textdegree, allowing cancellation of the second order contributions. In the interleaved case, with sextupoles placed every 90\textdegree, the second sextupole now makes a kick which prevents the cancellation and feeds into higher orders. So, unlike in the non-interleaved case, sextupole contributions continue downstream.}
\label{interleaved}
\end{figure*}

The potential usefulness of this feature can be explained using an example of sextupoles acting in one plane. At second order in sextupole strengths, cross terms between sextupoles appear. A non-interleaved scheme has an advantage in a 90\textdegree\ FODO section, that a contribution from sextupole can be countered by the next sextupole which is 180\textdegree away, this is not the case with a sextupole in every cell, this no longer occurs, enabling a pathway for growing higher order terms, modeled in Fig \ref{interleaved}..

\subsection{Test Lattice}

To isolate the effects of different correction schemes, a simplified test lattice was constructed based on the ESR optics. The test lattice was based on the EIC's version 5.3 ESR. and uses the same FODO optics as the 18GeV ESR and a single IP. An additional short straight section was placed opposite the IP to keep the geometry consistent. The Interaction Region (IR) uses the same $\beta^*$ and quadrupole placement. The 90\textdegree\  FODO's from the ESR are used, with the same number of sextupoles. The fractional tune is kept at 0.119 in the horizontal plane and 0.104 in the vertical. 

\begin{figure}[ht]
\centering
\includegraphics[width=\linewidth]{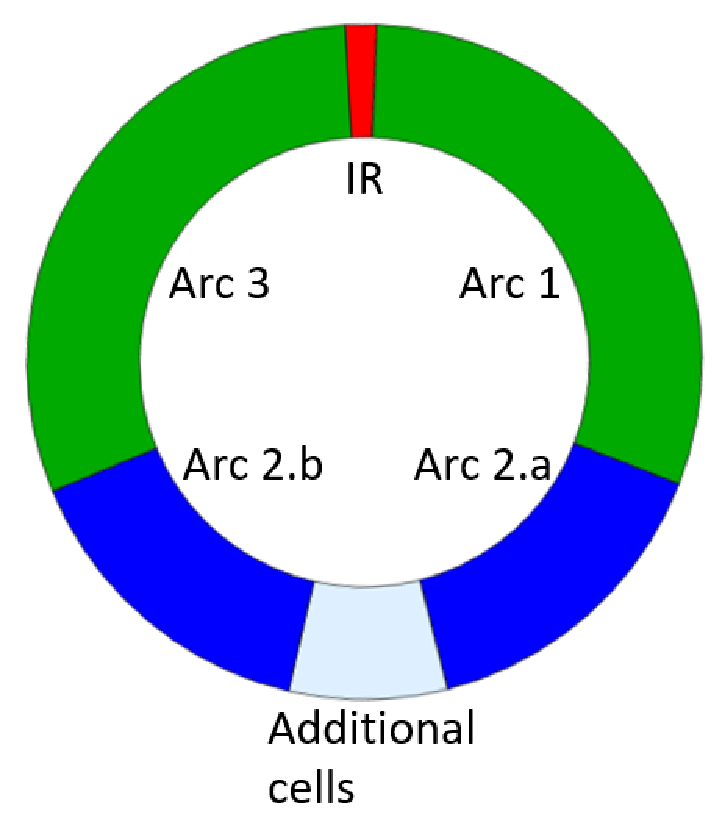}
\caption{\label{test_lattice} Schematic view of test ring. The W-function is 0 at the IP and is excited by the downstream IR; it is brought down over arc 1. Arc 2 is used for additional chromaticity correction. Additional cells to close the ring are added to the center of arc 2, splitting it into two parts. These additional arcs do not have sextupoles. Arc 3 mirrors arc 1, bringing the W-function back up to be compensated back to 0 by the upstream IR.}
\end{figure}

The test lattice had three arcs, the arcs on either side of the IR (arcs 1 and 3) are used for W-function correction and arc 2 is used for additional chromaticity correction. All three arcs have the same number of sextupoles, however, in order to make sure the lattice had a closed geometry, additional cells without sextupoles were placed in arc 2 (Fig.~\ref{test_lattice}). Arcs 1 and 3 share the sextupole families SF, SD, WF, and WD. These follow the convention of weak and strong sextupoles of the ATS scheme. Arc 2 has two sextupole families, SX1 and SX2.

For the purpose of this test, two chromatic effects will be corrected, the chromaticity and the W-function. The chromaticity will be set to one and the W-function will be set to zero at the IP and brought down over the first arc.

Lattice optimizations and dynamic aperture calculations were accomplished using the Bmad \cite{Bmad} library, with dynamic aperture calculations being ran for 2000 turns, which is roughly double the damping time for the ESR. Particle tracking used Bmad's 6D standard tracking.

\subsection{Partially interleaved 2-Family scheme}

As described previously, the partially interleaved 2-family scheme (shortened to 2-family scheme) uses the sextupoles designated as strong, leading to the chromatic conditions for correcting the IR being

\begin{subequations}
\begin{align}
\begin{split}
     N_{Sx}\beta_{x,Sx}D_{x,Sx}k_{Sx} \\ +N_{Sy}\beta_{x,Sy}D_{x,Sy}k_{Sy} =I_{Q_x}-1 \label{eqn: chrom a}\\
\end{split}
\\
\begin{split}     
N_{Sx}\beta_{y,Sx}D_{x,Sx}k_{Sx} \\ +N_{Sy}\beta_{y,Sy}D_{x,Sy}k_{Sy}=I_{Q_y}-1\label{eqn: chrom b}
\end{split}
\end{align}
\end{subequations}

\begin{subequations}
\begin{align}
\begin{split}
    e^{-2i(\Delta\mu_x-\pi/2)}[N_{Sx}\beta_{x,Sx}D_{x,Sx}k_{Sx}\\
    +e^{-i\pi/2}N_{Sy}\beta_{x,Sy}D_{x,Sy}k_{Sy}]=I_{W_x} 
\end{split}   
\\
\begin{split}
    e^{-2i(\Delta\mu_y-\pi/2)}[e^{-i\pi/2}N_{Sx}\beta_{y,Sx}D_{x,Sx}k_{Sx}\\
    +N_{Sy}\beta_{y,Sy}D_{x,Sy}k_{Sy}]=I_{W_y}
\end{split} 
\end{align}
\end{subequations}

\begin{figure}[ht]
    \centering
    \includegraphics[width=1\linewidth]{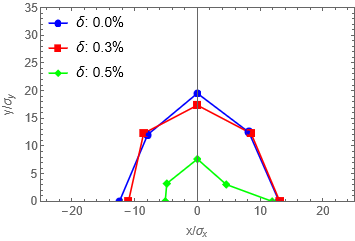}
    \caption{Test lattice dynamic aperture after W-function and chromaticity correction using two sextupole families in the W-function correction arcs.}
    \label{testDA2fam}
\end{figure}

\noindent Where $I_{Q_{x,y}}$ and $I_{W_{x,y}}$ are the contributions from the IR in each plane, $N_{Sx,Sy}$ are the number of sextupoles, $\Delta\mu_{x,y}$ are the phase advances from the IP. The chromaticity is corrected to 1 in order to avoid the head-tail instability \cite{}. These equations represent six conditions and cannot all be corrected by the two sextupole families and phase advances available. This issue is solved by having the W-function corrected by the arcs neighboring the IR and correcting the remaining chromaticity in arc 2 using the families SX1 and SX2. This scheme lead to DA seen in Fig. \ref{testDA2fam}.

\subsection{ATS scheme}

\begin{figure}[htbp]
    \centering

    \begin{subfigure}{1\columnwidth}
        \centering
        \includegraphics[width=0.9\linewidth]{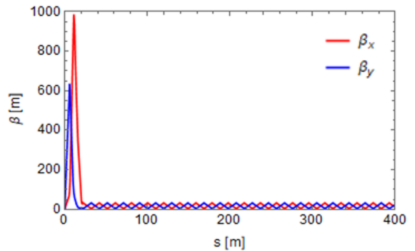}
        \caption{Periodic FODO cells in arc.}
        \label{fig:top}
    \end{subfigure}

    \begin{subfigure}{0.9\columnwidth}
        \centering
        \includegraphics[width=\linewidth]{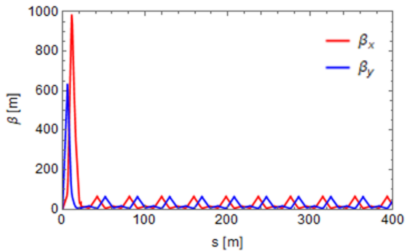}
        \caption{ATS optics in arc.}
        \label{fig:bottom}
    \end{subfigure}

    \caption{$\beta$-functions for the downstream IP going into the arcs for the test lattice for periodic FODO optics (a) and ATS optics (b).}
    \label{ATSOptics}
\end{figure}

\begin{figure}[ht]
    \centering
    \includegraphics[width=1\linewidth]{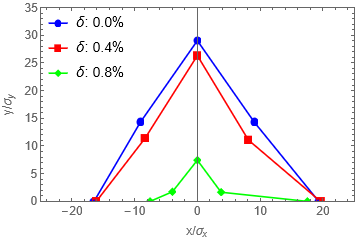}
    \caption{Test lattice dynamic aperture after W-function and chromaticity correction with the ATS scheme.}
    \label{testDAATS}
\end{figure}

The sextupoles for the ATS scheme were chosen using the same conditions as the 2-family scheme, resulting in different sextupole strengths due to the now present $\beta$-beat, for now keeping the weak sextupoles turned off. For this test, a telescopic ratio of 2 was used, leading to the optics in Fig \ref{ATSOptics}. The momentum and vertical aperture were substantially increased compared to the 2-family scheme, shown in Fig, \ref{testDAATS}. The cross talk of sextupoles acting in different planes is also reduced, shown in table \ref{beta_table} where the ratio of $\beta$-functions has decreased between planes.

\begin{table}[ht]
\centering
\caption{\label{beta_table} The $\beta_x$ at the SF and SD families are shown. The ratio of $\beta_x$ at the horizontal family to $\beta_x$ at the vertical family is seen to improve when using the ATS optics, giving a less interleaved scheme}
\begin{tabular}{|c|c|c|c|}
     \hline
     & $\beta_{x,SF}$[m] & $\beta_{x,SD}$[m] & $\beta_{x,SF}/\beta_{x,SD}$\\
     \hline
     2 and 4 family & 29.03 & 6.68 & 4.35 \\
     \hline
     ATS & 57.86 & 8.38 & 6.90 \\
     \hline
\end{tabular}

\end{table}

\subsection{4-Family scheme}

The 4-family scheme adds two new parameters, more than are fixed by the current conditions. A choice to fix the sextupoles of arc 2 to match those of the ATS scheme was made so that each arc is still responsible for the same portion of the correction. This choice of sextupole strength for arc 2 makes the changes between the two schemes only in arcs 1 and 3, also making it a more direct comparison between the two schemes.

\begin{figure}[ht]
    \centering
    \includegraphics[width=1\linewidth]{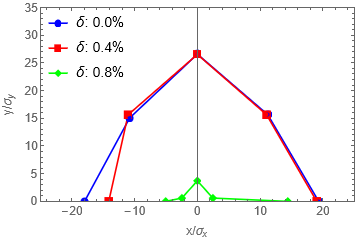}
    \caption{Test lattice dynamic aperture after W-function and chromaticity correction with four sextupole families for the arcs neighboring the IP.}
    \label{testDA4fam}
\end{figure}

The dynamic aperture calculations in the test lattice are without RF oscillations, as no cavities are present. Of the three schemes, the 2-family scheme performed significantly worse than the others at 0.5\%, with the ATS and 4-family performing similarly at 0.8\% shown in Fig. \ref{testDA4fam}. The ATS scheme had slightly better performance at larger momentum offsets, and the 4-family scheme performed slightly better on-momentum. 

\section{Pre-excited sextupoles in the test lattice}

In the prior sections, the ATS scheme used solely the strong sextupole families, however, use of the weak sextupoles may improve performance if the strengths are well chosen. For this test, it was chosen to initially set both the strong and weak sextupole families to correct the linear chromaticity of the arc. This choice makes the phase advance of each cell first order momentum independent, giving a clean initial setup. On top of this initial sextupole setting, the schemes in the previous section were applied. This results in lower sextupole strengths in the chromaticity correction arc.

\begin{figure}[ht]
    \centering
    \includegraphics[width=1\linewidth]{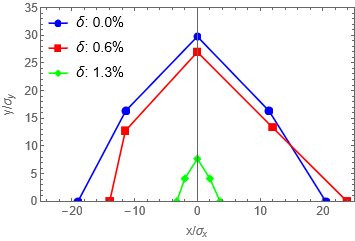}
    \caption{Test lattice dynamic aperture after W-function and chromaticity correction with four sextupole families and pre-excited sextupoles in the arcs neighboring the IP.}
    \label{testDA4famPre}
\end{figure}

\begin{figure}[ht]
    \centering
    \includegraphics[width=1\linewidth]{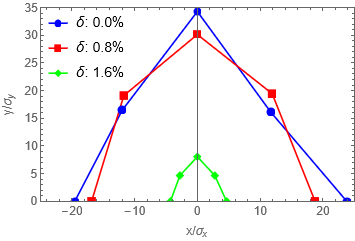}
    \caption{Test lattice dynamic aperture after W-function and chromaticity correction with with ATS scheme and pre-excited sextupoles.}
    \label{testDAATSPre}
\end{figure}

With the pre-excited sextupoles, all three schemes saw an improvement in performance. The 2-family scheme was increased to 0.9\%, and the ATS scheme both exceeded 1\%, at 1.6\% and 1.3\% respectively (Figs. \ref{testDA4famPre} and \ref{testDAATSPre}). These results give an indication that the ATS scheme may be beneficial for increasing the momentum aperture.

\section{ATS in the ESR}

An implementation of the three schemes can also be done for the full ESR lattice. The 5.3 1-IP ESR lattice was used for this test, with the matching for the ATS scheme being achieved with matrix elements. This solution was a highly optimized version of a 4-family per arc scheme described previously. For this test, the RF cavities were turned off, this was in order to remain comparable to the test lattice which did not have RF cavities. 

The ESR has six arcs, as opposed to the test lattices three. This gives more flexibility in application of the W-function reduction portion of the three schemes. The W-function can be corrected over one or two arcs. Both of these methods can be used, with the two arc method assigning sextupoles to the same task as in the test lattice. For comparison purposes, the 4-family scheme used here is the simplified version described in the test lattice, not more complicated baseline scheme described previously. This is done so that all correction schemes are compared with direct correction only of the linear chromaticity and W-function which are the relevant quantities for the altered arcs. 

Both of the methods were used for all three schemes with pre-excited sextupoles. Based on the test lattice, it would be reasonable that using the pre-excited sextupole variation would give better performance in the ESR. This was checked, and the pre-excited sextupole variations performed better in all cases, so the results below focus on the schemes with pre-excited sextupoles.

For both the one arc and two arc cases, the ATS scheme had the best performance, achieving 0.6\% compared to the 0.4\% and 0.5\% of the other schemes. The two arc scheme did see a reduction in the on-momentum aperture for the three schemes.

\begin{table*}[ht]
    \centering
    \caption{ESR summary. ATS performs best in the mometnum aperture, and ties with the 1-arc 2-family for best transverse aperture. Note that max sextupole strengths are not a good predictor of performance for these schemes.}
    \resizebox{\linewidth}{!}{
    \begin{tabular}{|c|c|c|c|c|}
        \hline
         & Momentum aperture & \makecell{Transverse \\aperture } & \makecell{Largest sextupole W-function \\ correction arcs [m\textsuperscript{-3}]} &  \makecell{Largest sextupole \\chromaticity correction arcs [m\textsuperscript{-3}]}\\
         \hline
         1-arc 2-family & 0.5\% & 16 & 5.01 & 3.13\\
         \hline
         1-arc ATS & 0.6\% & 16 & 3.36 & 3.00\\
         \hline
         1-arc 4-family & 0.5\% & 13 & 5.67 & 3.00\\
         \hline
         2-arc 2-family & 0.5\% & 12 & 3.51 & 4.99\\
         \hline
         2-arc ATS & 0.6\% & 11 & 2.22 & 5.14\\
         \hline
         2-arc 4-family & 0.4\% & 10 & 2.82 & 5.14\\
         \hline
    \end{tabular}
    }
    
    \label{ESR_results}
\end{table*}

\begin{figure}[ht]
    \centering
    \includegraphics[width=\linewidth]{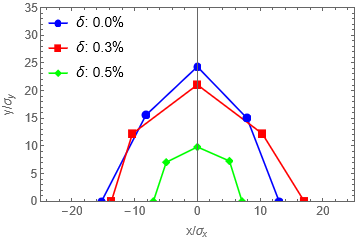}
    \caption{ESR dynamic aperture after W-function and chromaticity correction with 4-family scheme and pre-excited sextupoles. With W-function corrected over the IP's neighboring arcs.}
    \label{ESRDA4fampre}
\end{figure}

\begin{figure}[ht]
    \centering
    \includegraphics[width=\linewidth]{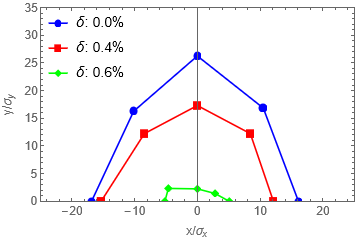}
    \caption{ESR dynamic aperture after W-function and chromaticity correction with with ATS scheme and pre-excited sextupoles. With W-function corrected over the IP's neighboring arcs.}
    \label{ESRDAATSpre}
\end{figure}

The dynamic aperture results from the ATS scheme performed favorably when compared to the 4-family scheme, increasing the momentum aperture by 0.1\% before the pre-excited sextupoles are used. After the implementation of the pre-excited sextupoles, the momentum aperture is increased by another 0.1\% over the 4-family scheme, resulting in a 0.2\% advantage in the ATS scheme for this test, shown in Figs. \ref{ESRDA4fampre} and \ref{ESRDAATSpre}.

Across all configurations studied, the ATS scheme achieves the largest momentum aperture, improving momentum aperture by approximately 0.1–0.2\% relative to conventional schemes.

As seen, the ATS scheme applied to the ESR offers a small benefit to the momentum aperture over the competing schemes. However, this improvement is accompanied by an increase in emittance due to the altered optics in the arcs, which may offset the benefits for applications requiring low emittance.\cite{Unger:2022zsb}. 

The increase of the $\beta$-function used in the ATS scheme is accompanied by an increase of the emittance. The ESR lattice used started with an emittance of 29.6nm. Applying the ATS scheme over one arc saw an increase of 5.3\% and over two arcs saw an increase of 10\%. The increase can be seen by looking at the first order emittance calculation

\begin{equation}
    \epsilon_a=\frac{C_q}{I_2-I_{4a}}\gamma_0^2 I_{5a}
    \label{lin_emit}
\end{equation}

\noindent where the $I$s are the standard radiation integrals, $\gamma_0$ is the relativistic factor for the beam, and $C_q$ is a particle species specific constant

\begin{equation}
    C_q=\frac{55}{32\sqrt{3}}\frac{\hbar}{mc}
\end{equation}

The only quantity in  that changes in eq. \ref{lin_emit} under the ATS scheme is $I_{5a}$ which is

\begin{equation}
    I_{5a}=\oint \frac{1}{\rho^3}\mathcal{H}_a ds
\end{equation}

\noindent where $\rho$ is the radius of curvature and $\mathcal{H}_a$ is

\begin{equation}
    \mathcal{H}_a=\gamma_a \eta_a^2 + 2 \alpha_a \eta_a \eta_a' + \beta_a \eta_a'^2
\end{equation}

It can be seen that the change in emittance from this calculation is dependent on the particulars of the lattice such as dipole placement which determines where the optics change is important. In the ESR, the dipoles take up most of the space between quadrupoles, so what matters here is the change in $\mathcal{H}_a$. An approximation of the emittance increase can be found under the assumption that the average $ \mathcal{H}_a$ over the cell is the driving factor.

with the average being change in $\beta$-function in the arc being

\begin{equation}
    \langle B \rangle=\frac{1}{2}\left(T_r+\frac{1}{T_r}\right)
\end{equation}

And its inverse for $\gamma$

\begin{equation}
    \langle \frac{1}{B} \rangle=\frac{1}{2}\left(T_r+\frac{1}{T_r}\right)
\end{equation}

Assuming approximate canceling from oscillations in the $\alpha$ term of $ \mathcal{H}_a$, the average then changes to

\begin{equation}
     \langle \mathcal{H}_a \rangle \approx \gamma_a \eta^2 \langle \frac{1}{B} \rangle+2 \alpha_a \eta_a \eta_a' + \beta_a \eta_a'^2 \langle B \rangle
\end{equation}

\noindent which, under the previous assumptions, becomes

\begin{equation}
     \langle \mathcal{H}_a \rangle \approx \frac{1}{2}\left(T_r+\frac{1}{T_r}\right)\mathcal{H}_{a,0}
\end{equation}

For the telescopic ratio of 2 used in this study, this amounts to an increase in emittance of the arc of 25\%, which gives an estimated increase of overall emittance change of 8\% under the assumption that all emittance is created in the arcs, which is in rough agreement to the change seen.

In addition, the increased beam size in the arcs would require an increase in magnet aperture for the arcs where the ATS scheme is applied. A larger telescopic ratio could increase the ATS scheme's performance, however this would also increase the mentioned issues. For the ESR, where emittance constraints are stringent, this trade-off limits the practical applicability of ATS despite its benefits for momentum aperture.

This shows that when schemes are compared that are based on linear chromaticity and W-function, the ATS scheme produces a small advantage in momentum aperture in comparison to the other schemes. However, the emittance increase leaves the ATS scheme unsuitable for the ESR. It does, however, become a possible scheme for other ring accelerators where an emittance increase from the arcs is acceptable.

\section{conclusion}
We have evaluated the application of the Achromatic Telescopic Squeezing scheme to the ESR of the EIC, comparing its performance to conventional sextupole correction strategies. Using both a test lattice and the full ESR lattice, we find that ATS reduces sextupole strengths and improves momentum aperture through more favorable nonlinear dynamics.

Without the use of pre-excited sextupoles, DA results were similar to standard optics, giving a possible scheme with a  reduced number of sextupoles in those arcs. When included, the pre-excited sextupoles further enhanced performance across all schemes when chosen appropriately, with ATS achieving the largest momentum acceptance. However, the altered optics inherent to ATS lead to emittance growth, which limits its suitability for the ESR.

These results indicate that while ATS is not optimal for the ESR, it remains a promising approach for storage rings where moderate emittance growth is acceptable and improved momentum aperture is needed or a reduction in sextupole strength or number is desirable.

\bibliography{apssamp}

\end{document}